
\vskip 24pt
\magnification=\magstep1
\hsize=6truein
\vskip 20pt
\font\tfont=cmbx12 scaled\magstep2
\font\eightrm=cmr8

\leftskip=.25in
\centerline\tfont{\bf SINGLE AND DOUBLE BFKL POMERON EXCHANGE}

\centerline{\bf AND A DIPOLE PICTURE
OF}
\centerline{\bf HIGH ENERGY HARD PROCESSES}
\vskip 20pt
\centerline{A.H. Mueller and Bimal Patel}
\vskip 6pt
\centerline{Department of Physics, Columbia University}
\centerline{New York, New York 10027}
\vskip 10pt
\def\today{\ifcase\month\or January\or February\or March\or \April\or \May\or
June\or July\or August\or September\or October\or November\or December\fi
\space \number \day, \number\year}{\centerline\today}

\vskip 30pt

{\narrower\smallskip
Onium-onium scattering at high energy is used to illustrate
a dipole picture of high energy hard scattering in the large $N_c$ limit.
Single and double BFKL pomeron exchanges are calculated in the leading
logarithmic approximation.  An expression is given for the triple pomeron
coupling
when one of the pomeron's momentum is zero while the other two have momentum
transfer, t.  This expression is explicit and could be evaluated numerically.
It
has a $(-t)^{-{1\over 2}}$ singularity at t=0.
\smallskip}

\vskip 20pt
\baselineskip=18pt
\centerline{\bf 1.  Introduction}

\indent The Balitsky, Fadin, Kuraev and Lipatov (BFKL)[1-3] pomeron is one of
the most intriguing objects in perturbative QCD.  It applies to processes which
are at the same time hard processes and high energy processes.  Although in
principle it is straightforward to measure experimentally[4-8], in practice the
necessary data are hard to get.  On the conceptual level perhaps the simplest
process where the BFKL pomeron applies is in very high energy onium-onium
scattering.  For a sufficiently heavy onium state high energy onium-onium
scattering is a perturbative process since the onium radius gives the essential
scale at which the running coupling is evaluated.  For $s/M^2$, with  M  the
onium mass, large but not too large it is a good approximation to neglect the
running of the coupling completely in which case the onium-onium cross section
behaves as $\sigma \sim exp\{(\alpha_P-1) \ln s/M^2\}$ with
$\alpha_P={4\alpha C_A\over \pi}\ln 2.$  The growth of the cross section
violates unitarity bounds when $(\alpha_P-1) \ln s/M^2$ becomes large.  When
$M^2$ is large enough this violation of unitarity occurs in a regime where the
fixed coupling approximation is valid.  Multi-pomeron exchanges should slow the
rate of growth of the cross section to keep it consistent with unitarity.

\indent From the partonic point of view the growth of the cross section in
onium-onium scattering has a simple interpretation.  A high energy onium state
consists of a heavy quark-antiquark pair and a large number of  soft gluons.
In the large $N_c$ limit this system can be viewed as a collection of color
dipoles[9,10].  Then in the center of mass of a high energy onium-onium
scattering the cross section can be understood as a product of the number of
dipoles in one onium state times the number of dipoles in the other onium state
times the basic (energy independent) cross section for dipole-dipole scattering
due to two gluon exchange.  The cross section grows rapidly with energy because
the number of dipoles in the light-cone wavefunction grows rapidly with
energy.  When that number becomes sufficiently large the single scattering
aproximation between dipoles in the colliding onia ceases to become valid and
double and higher number of scatterings may become important.

 In this paper, we make this dipole picture of high energy scattering
explicit by calculating onium-onium scattering in the way  described above.
The result here is not new[1].  We also carry out the double dipole
scattering,  the two pomeron exchange term.  This result is new and is given
by (5), (8), (43) and (51).

\indent We also calculate the triple pomeron coupling, still in the large $Nc$
limit and in the leading logarithmnic approximation.  As with the two
pomeron contribution discussed above our result, eq.(61), depends on the
quantity ${\rm V}_\nu$, given by (43).  We have not been able to get an
analytic
expression for $V_\nu$, or even for ${\rm V}_0$ which appears in (61).
However,
(43) is easily invertible for ${\rm V}_\nu$ and a numerical evaluation of V may
be possible.

\indent Our discussion in this paper has been restricted to onium-onium
scattering.  As regards the triple pomeron coupling we view the use of
onium-onium scattering as a device to obtain this more universal quantity.
There are known physical processes where a single BFKL
pomeron exchange gives the correct physics[4-9].  We expect that the discussion
given here could be repeated for those processes.

\vskip 20pt
\centerline{\bf 2. Single Pomeron Exchange in Onium-Onium Scattering}

\indent In this section we consider onium-onium scattering in the leading
logarithmic approximation.  The discussion here is a slight extension of
ref.[9], and as in that reference we find it convenient to use
$\psi_{\alpha\beta}^{(0)}(\b k_1,z_1)$, the onium light-cone
wavefunction\footnote{**}{\eightrm An approach similar to that
given in [9] has recently been given by N.N. Nikolaev, B.G. Zakharov and V.R.
Zoller,
KFA-IKP preprint (January 1994) and by N.N. Nikolaev and B.G. Zakharov KFA-IKP
preprint (January 1994).  I wish to thank E.M. Levin for bringing this work to
my
attention.  These authors also emphasize a dipole picture of high energy hard
scattering.}
with no soft gluons.
$\alpha$ and
$\beta$ are the spin indices of the quark and antiquark respectively while $\b
k_1$
and
$z_1\  p_+$ are the transverse and light-cone momenta of the antiquark with  p
the
onium momentum.  We suppose
$\b p = 0.$  The wavefunction

$$\psi_{\alpha\beta}^{(0)} (\b x_{01}, z_1) = \int {d^2k_1\over (2\pi)^2}
e^{i\b k_1\cdot \b x_{01}}\psi_{\alpha\beta}^{(0)}(\b k_1, z_1)\eqno(1)$$

\noindent is in a mixed representation.  We suppose the quark has transverse
coordinate $\b x_0$ and the antiquark transverse coordinate $\b x_1$,  and we
define $\b x_{01} = \b x_1-\b x_0.$  It is also useful to define

$$\Phi^{(0)}(\b k, z) = \Sigma_{\alpha\beta}|\psi_{\alpha\beta}^{(0)}(\b k,
z)\vert^2\eqno(2)$$
\noindent and

$$\Phi^{(0)}(\b x, z) = \Sigma_{\alpha\beta}|\psi_{\alpha\beta}^{(0)}(\b x,
z)|^2\eqno(3)$$

\noindent where

$$\int {d^2\b k\over (2\pi)^2} \int_0^1 dz\  \Phi^{(0)}(\b k,  z) = \int d^2 \b
x
\int_0^1 dz\  \Phi^{(0)}(\b x, z) = 1.\eqno(4)$$
\bigskip

\noindent{2.1  THE TWO GLUON EXCHANGE APPROXIMATION}

To set our normalization we now evaluate the forward onium-onium scattering
amplitude in the two gluon approximation.  The process is illustrated in Fig.1
where one of
the relevant graphs is shown.  We suppose $p_+=p_-^\prime$ are the large
momenta with $\b p
=\b p^\prime = 0.$  Let  A  be the scattering amplitude normalized according to

$${d\sigma\over dt} = {1\over 4\pi} |A|^2.\eqno(5)$$

\noindent  A can be evaluated in terms of the scattering of the free
quark-antiquark pair
$(p-k_1,k_1)$ on the free quark-antiquark pair $(p^\prime -
k_1^\prime,k_1^\prime)$
since the time of interaction between the two systems is very short compared to
the
time scales in the onium wavefunctions.  For example, the sum of all graphs
where the
scattering is only between quark
$k_1$ and quark $k_1^\prime$ is given by

$$A_{qq} = - i \int {d^2k_1\over (2\pi)^2} \int_0^1 dz_1 \int
{d^2k_1^\prime\over
(2\pi)^2}
\int_0^1 dz_1^\prime \Phi(\b k_1,z_1) \Phi(\b k_1^\prime,z_1^\prime) a
\eqno(6)$$
with $a$ the forward quark-quark scattering amplitude given
by

$$a = {-1\over 2} \alpha^2 \int {d^2 \ell\over [\b{$\ell$}^2]^2}, \eqno(7)$$

\noindent where the color factor ${N_c^2-1\over 4N_c^2}$ has been set equal to
1/4. $a$
is infrared divergent but this divergence will disappear when the other graphs
are
included.  The final answer is most compactly given in transverse coordinate
space as

$$A  = - i \int d^2x_{01} d^2x_{01}^\prime \int_0^1 dz_1 dz_1^\prime \Phi (\b
x_{01},  z_1) \Phi (\b x_{01}^\prime,  z_1^\prime) F\eqno(8)$$

\noindent with

$$F =  - {\alpha^2\over 2} \int {d^2\ell\over [\b{$\ell$}^2]^2} (2-e^{-i\b
{$\ell$}\cdot \b
x_{01}}-e^{i\b{$\ell$}\cdot \b x_{01}})(2-e^{-i \b{$\ell$}\cdot \b
x_{01}^\prime} -
e^{i\b
{$\ell$}\cdot
\b x_{01}^\prime}).\eqno(9)$$

\noindent (Eqs.(7) and (9) are derived in the Appendix.)  The infrared
divergence is
now removed.
\bigskip
\noindent{2.2.  THE DIPOLE DENSITY IN AN ONIUM STATE}

\indent Eqs.(8) and (9) express the onium-onium scattering amplitude as the
scattering of
two dipoles each made of the heavy quark-antiquark pair comprising an onium
wavefunction.
In the large $N_c$ limit and in the leading logarithmnic approximation the
complete,
multi-gluon, wavefunction of an onium state can be viewed as a collection of
dipoles since
each gluon acts like a quark-antiquark pair.  In this section we calculate the
density of
such dipoles in the onium wavefunction.  In the next section that dipole
density
will be used to calculate high energy onium-onium scattering in a very physical
way.

\indent We define $n(x_{10}, x,Y)$ such that

$$N(x,Y) = \int d^2x_{01} \int_0^1 dz_1 \Phi (x_{01},z_1) n(x_{01}, x,
Y)\eqno(10)$$

\noindent is the number density of dipoles of transverse coordinate separation
x  with the
smallest light-cone momentum in the pair greater than or equal to $e^{-Y}p_+$
where
$p_+$ is the light-cone momentum of the onium.  then $\int_0^\infty {dx\over x}
N(x, Y)$ gives the total number of dipoles in the onium state satisfying the
longitudinal momentum constraint given above.  n  obeys the integral equation

$$n(x_{01},x,Y) = exp\left\{- {4\alpha C_F\over \pi} \ln
\left(x_{01}/\rho\right)Y\right\} x \delta(x-x_{01}) + {4\alpha C_F\over \pi}
\int_0^Y dy$$
$$\cdot  exp\left\{-{4\alpha C_F\over \pi} \ln
\left(x_{01}/\rho\right)(Y-y)\right\}\tilde K(x_{01}, x_{12}) dx_{12}
n(x_{12},x,y).\eqno(11)$$

\noindent This equation is identical in form to that of eq.(24) of ref[9]
except
for the inhogeneous term which reflects the fact that the dipole can be formed
from the valence heavy quark-antiquark pair without any soft gluons in the
wavefunction.  The kernel $\tilde K$ is given by  eq.(25) of ref[9], and we
remind
the
reader that $\rho$ is an ultraviolet cutoff which will soon  be set to zero.
Eq.(11)
is schematically illustrated in Fig.2 where the double line, labelled by 2, is
a
gluon resolved into its quark and antiquark components in the large $N_c$
approximation.

\indent Eq.(11) is easily solved by writing

$$n(x_{01}, x, Y) = \int {d\omega\over 2\pi i} e^{\omega Y} n_\omega (x_{01},
x)\eqno(12)$$

\noindent where the $\omega$ integral goes parallel to the imaginary axis and
to
the right of any $\omega$-singularities in $n_\omega.$  Using (12) in (11) one
finds

$$n_\omega (x_{01},x) = {x\delta(x_{01}-x)\over\omega}+ {4\alpha C_F\over \pi
\omega}
\int dx_{12} K(x_{01}, x_{12}) n_\omega (x_{12}, x)\eqno(13)$$

\noindent where

$$K(x_{01}, x_{12}) = \tilde K(x_{01},x_{12}) - \delta(x_{01}-x_{12}) \ln
(x_{01}/\rho)\eqno(14)$$

\noindent is the BFKL kernel in the limit $\rho \to 0.$  to solve (13) write

$$n_\omega (x_{10}, x) = \int {d\nu\over 2\pi} (x_{01}/x)^{1+2i\nu}
n_{\nu\omega}\eqno(15)$$

\noindent where the $\nu$-integral goes along the real axis.  Using

$$\int dx_{12} K(x_{01}, x_{12}) x_{12}^{1+2i\nu} = \chi (\nu)
x_{10}^{1+2i\nu}\eqno(16)$$

\noindent with

$$\chi (\nu) = \psi(1) - {1\over 2}\psi({1\over 2} + i \nu) - {1\over 2} \psi
({1\over 2} -i\nu)\eqno(17)$$

\noindent where $\psi (x) = \Gamma^\prime(x)/\Gamma (x),$  we find from (13)

$$n_{\nu\omega} = {2\over \omega - {4\alpha C_F\over \pi} \chi
(\nu)}.\eqno(18)$$
Using (18) it is easy to determine\ $n(x_{01}, x, Y)$ as

$$n(x_{01}, x, Y) = {1\over 2} (x_{01}/x) {e^{(\alpha_P-1)Y}\over
{\sqrt{7\alpha
C_F \zeta (3)Y}}} exp\{-{\pi \ln^2 x_{01}/x\over 28 \alpha C_F \zeta (3)
Y}\}\eqno(19)$$

\noindent in the saddle point approximation in $\nu$ so long as $\ln\
x_{01}/x <<
\alpha C_F Y.\ \   \alpha_P-1\break
={8\alpha C_F \ln 2\over \pi}.$

\bigskip
\noindent{2.3.  SINGLE POMERON EXCHANGE FOR ONIUM-ONIUM SCATTERING IN TERMS OF
DIPOLE-DIPOLE SCATTERING}
\bigskip
\indent We now consider onium-onium scattering.  We may write the onium-onium
forward scattering amplitude as in (8), but where  F  is now given by

$$F = - {\alpha^2\over 2} \int {d^2\ell\over [\b{$\ell$}^2]^2}n(x_{01},x,Y/2) n
(x_{01}^\prime, x^\prime, Y/2) {d^2 x^\prime d^2x\over 4\pi^2x^2 x^{\prime
2}}$$

$$\cdot (2-e^{-i\b{$\ell$} \cdot \b x}-e^{i\b{$\ell$}\cdot \b x})(2-e^{-i\ell
\cdot
\b
x^\prime} -e^{i\b{$\ell$} \cdot \b x^\prime}).\eqno(20)$$

 Eq.(20) expresses the onium-onium forward scattering amplitude in terms
of the product of the dipole number densities in each of the onium states times
the
dipole-dipole scattering amplitude given by (19). We view the process in the
center
of mass system so that the Y/2 argument in the n's in (20) reflects the
requirement
that the dipoles which partake in the scattering be moving in the same
direction as
the onia of which they are (respectively) a part.  We have let our normal
integration factor dx/x become ${d^2\b x\over 2\pi x^2}$ in order to take into
account the fact that the dipole-dipole scattering is dependent on the
orientation
of the dipoles.  n  does not depend on the dipole orientation, however.

\indent After evaluating the angular integrals in (20) and using (19) we arrive
at

$$F={-2\pi\alpha^2x_{01}x_{01}^\prime e^{(\alpha_P-1)Y}\over 7\alpha C_F\zeta
(3)Y}I\eqno(21)$$

\noindent where

$$I=\int_0^\infty{dx\over x^2} {dx^\prime\over x^{\prime 2}}{d\ell\over
\ell^3}(1-J_0(\ell x))(1-J_0(\ell x^\prime)) exp\{-a(\ln^2 x_{01}/x + \ln^2
x_{01}^\prime/x^\prime)\}\eqno(22)$$

\noindent with

$$a = a (Y) = [14 {\alpha C_F\over \pi} \zeta (3)Y]^{-1}.\eqno(23)$$

\noindent  I  is easily evaluated
by rescaling  x  and  ${\rm x}^\prime$ so that

$$ I = \int_0^\infty {du\over u^2} {dv\over v^2} {d\ell\over \ell}
(1-J_0(u))(1-J_0(u)) exp\{-a(\ln^2 {x_{01}\ell\over u} + \ln^2{x_{01}^\prime
\ell\over v})\}.\eqno(24)$$

\noindent In the leading logarithmic approximation we can neglect  u  and  v
in
the exponential term on the right-hand side of (24).  Using

$$\int_0^\infty {du\over u^2} (1-J_0(u) = 1$$

\noindent we find

$$ I={\sqrt{7 \alpha C_F \zeta (3) Y}},\eqno(25)$$

\noindent so long as $\vert \ln(x_{01}^\prime/x_{01})\vert << a^{-1/2}.$
Thus,
the forward onium-onium scattering amplitude is given by (8) with

$$F = - 2\pi x_{01} x_{01}^\prime {e^{(\alpha_P-1)Y}\over {\sqrt{7 \alpha C_F
\zeta
(3) Y}}}.\eqno(26)$$

\noindent Eqs.(8) and (26) express the forward onium-onium scattering amplitude
in
terms of the BFKL pomeron.  Indeed, this process is similar to that originally
considered by Balitsky and Lipatov[1].  However, (20) shows that we may also
view
the process as the scattering of two dipoles in the onia by means of a 2-gluon
exchange.
\vfill
\eject
\centerline{\bf 3. Double Pomeron Exchange in Onium-Onium Scattering}

In this section, we generalize our discussion to include double pomeron
exchange in
high energy scattering.  In the dipole picture of onium-onium scattering two
pomeron exchange corresponds to the independent scatterings of two dipoles in
each
of the onia.  Our first task is to calculate the number density for a pair of
dipoles to be found in the light-cone  wavvefunction of an onium state.
\bigskip

\noindent{3.1  THE DIPOLE PAIR DENSITY IN AN ONIUM STATE}

In analogy with $n(x_{10}, x, Y)$ defined in (10) we introduce\break $n_2(\b
x_{01},
\b x_a, \b x_b, Y, \b q)$ as the dipole pair density in an onium state.  The
two
dipoles have transverse coordinate separation $\b x_a$ and $\b x_b,$
respectively.
Y  has exactly the same meaning as before and  q  is the transverse momentum
carried by each of the dipoles.  (That is, one of the dipoles has transverse
momentum $\b q$  while  the other has transverse momentum $-\b q$, with $q =
\vert \b
q \vert .)$  The equation which governs $n_2$ is

$$n_2(\b x_{01}, \b x_a, \b x_b, Y, \b q)
= {2\alpha C_F\over \pi^2} \int {x_{01}^2
d^2x_2\over x_{02}^2 x_{12}^2}
\int_0^Y$$
$$ exp\{{-4\alpha C_F\over \pi} \ln
({x_{01}\over \rho})(Y-y)\}dy\  e^{i\b q\cdot \b x_{01}/2}$$
$$\cdot n(\b x_{02}, \b x_a, y, \b q) n (\b x_{12}, \b x_b, y,\b q) + {2\alpha
C_F\over \pi^2}\ \int_0^Y dy\   exp\{{-4\alpha C_F\over \pi}\ln
({x_{01}\over
\rho})(Y-y)\}$$
$$\cdot \int {x_{01}^2 d^2x_2\over x_{02}^2 x_{12}^2} n_2 (\b x_{12}, \b x_a,\b
x_b
y,
\b q).\eqno(27)$$

\noindent The factor of 2 in front of the first term on the right-hand side of
(27)
accounts for an identical term with $\b x_a \leftrightarrow  \b x_b.$  The
process is
illustrated in Fig.3.  The factor $e^{i\b q\cdot \b x_{01}/2} = e^{i{\b q\over
2}[(\b x_0+\b x_2)-(\b x_1+\b x_2)]}$ corresponds to the dipole transverse
momentum
while $n(\b x, \b x^\prime, y,\b q)$ is defined as in sec.2.2 except that now
there is a nonzero transverse momentum, q, so that

$$n(\b x, \b x^\prime, y, \b q) = \int {d\omega\over 2\pi i} e^{\omega Y} \int
{d\nu\over 2\pi} (x/x^\prime){1\over 2} E_q^{0\nu *}(\b x^\prime) E_q^{0\nu}(\b
x)
n_{\nu\omega}\eqno(28)$$

\noindent with $n_{\nu\omega}$ as given in (18) and with $E_q^{0\nu}$ given in
ref.[3] as

$$E_q^{0\nu}(\b x) = {2\pi^2\over x b_\nu} \int d^2 \b R e^{i\b q\cdot \b R}
E^{0\nu}(\b R + \b x/2, \b R - \b x/2)\eqno(29)$$

\noindent where

$$E^{0\nu} (\b R + \b x/2,  \b R - \b x/2) = \biggl[{x\over \vert \b R + \b x/2
\vert \vert \b R - \b x/2\vert}\biggr]^{1+2i\nu}.\eqno(30)$$

\noindent $b_\nu$ is given in ref.[3] and has the property

$$b_\nu\mathrel{\mathop{\longrightarrow}\limits_{\nu \to 0}}-i\pi^3/\nu.$$

In order to solve (27) it is useful to have a convenient form for  n  for large
 y.
{}From (18) and (28) one finds

$$n(\b x, \b x^\prime, y, \b q) = {\pi\over 2} x\  e(\b x, \b q) {1\over
x^\prime}
e(\b x^\prime, \b q) {e^{(\alpha_P-1)y}exp\{-{a\over 2}\ln^2
x/x^\prime\}\over
[7 \alpha C_F \zeta(3) y]^{3/2}}\eqno(32)$$

\noindent for large  y  with a=a\ (y) given by (23) and where

$$e(\b x, \b q) = {1\over 2\pi} \int {d^2\b R e^{i\b q\cdot \b R}\over \vert \b
R -
\b x/2\vert \vert \b R + \b x/2\vert} .\eqno(33)$$

\noindent Using (32) in (27) and defining $n_2(x_{01}, Y, q)$ by

$$n_2(x_{01}, Y,q) exp\{-{a\over 2}(\ln^2  x_{01}/x_b + \ln^2
x_{01}/x_a)\}
{1\over x_a} e(\b x_a, \b q) {1\over x_b} e(\b x_b, \b q)$$
$$= \int_0^{2\pi}
{d\phi(\b x_{01})\over 2\pi} n_2(\b x_{01}, \b x_a, \b x_b,Y, \b q),\eqno(34)$$

\noindent with $\phi(\b x_{01})$ the angular orientation of $\b x_{01}$, one
finds

$$n_2(x_{01},Y,q) ={\pi x_{01}^2\over 8[\ln(x_{01}/\rho) + 4\ln 2]}
\int {d^2x_2\over x_{02} x_{12}} J_0(q x_{01}/2) e(\b x_{12}, \b q) e(\b
x_{02},
\b q)$$
$$  {e^{2(\alpha_P-y)Y}\over [7\alpha C_F\zeta(3)Y]^3}
+ {4\alpha C_F\over \pi} \int\tilde K(x_{01}, x_{12}) dx_{12} \int_0^Y dy\
exp\{-{4\alpha C_F\over \pi} \ln(x_{01}/\rho)(Y-y)\}$$
$$n_2(x_{12},
y,q).\eqno(35)$$

\noindent where we have used the fact that $x_{01}/x_{02}$ and $x_{01}/x_{12}$
do
not vary too far from 1 in order to cancel the ``diffusion'' terms in (35).
This is
correct so long as  q  is not too small.   Using

$${e^{2(\alpha_P-1)Y}\over Y^3} = - {1\over 2} \int {d\omega\over 2\pi i}
e^{\omega
Y}[2(\alpha_P -1)-\omega]^2\ln[\omega - 2(\alpha_P-1)]\eqno(36)$$

\noindent and defining

$$n_2(x_{01}, Y,q) = \int {d\omega\over 2\pi i} n_{2\omega} (x_{01}, q)
e^{\omega
Y}\eqno(37)$$

\noindent along with

$$I_\omega = {-\pi[2(\alpha_P-1)-\omega]^2\ln[\omega-2(\alpha_P-1)]\over
16[\ln(x_{01}/\rho) + 4 \ln 2][7 \alpha C_F\zeta(3)]^3}
\int {d^2x_2\over x_{02}x_{12}} J_0(q\b x_{01}/2) e(\b x_{12}, \b q) e(\b
x_{02}, \b
q)\eqno(38)$$

\noindent gives

$$n_{2\omega}(x_{01},q)=x_{01}^2  I_\omega+{4\alpha C_F/\pi\over \omega +
{4\alpha
C_F\over \pi}\ln(x_{01}/\rho)}\int
dx_{12}\tilde K(x_{01},x_{12})n_{2\omega}(x_{12},q).\eqno(39)$$

\noindent One can recast (39) to read

$$n_{2\omega}(x_{01},q)=x_{01}^2I_\omega\left[1+{4\alpha C_F\over
\pi\omega}\ln({x_{01}\over \rho})\right]+{4\alpha C_F\over \pi\omega}\int
dx_{12}K(x_{01},x_{12})n_{2\omega}(x_{12},q).\eqno(40)$$

\noindent Eq.(40) is easily solved by writing

$$n_{2\omega}(x_{01},q)=x_{01}^2\int {d\nu\over 2\pi} (q\
x_{01})^{-1+2i\nu}n_{2\nu\omega}.\eqno(41)$$

\noindent We arrive at

$$n_{2\nu\omega} = - {\pi[2(\alpha_P-1)-\omega]^2\ln[\omega -
2(\alpha_P-1)]V_\nu\over 16[4\ln 2-\chi(\nu)][7 \alpha
C_F\zeta(3)]^3}\eqno(42)$$

\noindent where

$$\int {d^2x_2\over x_{02}x_{12}}J_0(q x_{01}/2) e(\b x_{12},\b q) e(\b x_{02},
\b
q) =
\int {d\nu\over 2\pi}(q x_{01})^{-1+2i\nu}V_\nu\eqno(43)$$

\noindent and where we have taken the limit $\rho \to 0$ in the first term on
the
right-hand side of (40).  Using (37), (41) and (42) one obtains

$$n_2(x_{01},Y,q)={\pi x_{01}^2 e^{2(\alpha_\rho-1)Y}\over 8(7 \alpha C_F
\zeta(3)Y)^3}\int{d\nu\over 2\pi}(q\  x_{01})^{-1+2i\nu}{V_\nu\over 4 \ln
2-\chi(\nu)}.\eqno(44)$$

\noindent Eqs.(34) and (44) give $n_2(\b x_{01}, \b x_a, \b x_b, Y, \b q),$
averaged
over angles of $\b x_{01}$, in the leading logarithmnic approximation so long
as
$\ln({1\over q x_{01}}) << {\sqrt{\alpha Y}}$.  (When $q x_{01}$ becomes too
small our replacement of $exp\{-{a\over 2}(\ln^2 x_{12}/x_a + \ln^2
x_{12}/x_b)\}$ by
$exp\{-{a\over 2}(\ln^2 x_{01}/x_a + \ln^2 x_{01}/x_b)\}$ in using (34)
in (27)
is not reliable.)

\bigskip
\noindent{3.2.  THE DOUBLE SCATTERING CONTRIBUTION TO THE ONIUM-}
\noindent{ONIUM AMPLITUDE}

In this section we shall use the dipole pair density to calculate the
onium-onium
forward scattering amplitude when two pomerons are exchanged.  The exchange of
two pomerons corresponds to the double scattering approximation in terms of
dipoles
of one onium state scattering on dipoles of the other onium state.  The forward
scattering amplitude can be written as in (8) with  F  now given by

$$F={1\over 2!}\int {d^2q\over (2\pi)^2}n_2(\b x_{01}, \b x_a,\b x_b, Y/2,\b
q)n_2(\b x_{01}^\prime, \b x_a^\prime, \b x_b^\prime, Y/2,-\b q)$$
$$\cdot  {d^2x_ad^2x_a^\prime\over (2\pi x_ax_a^\prime)^2}{d^2x_b
d^2x_b^\prime\over (2\pi x_bx_b^\prime)^2}\{-{\alpha^2\over 2} {d^2\ell_a\over
\b
{$\ell$}_a^2(\b{$\ell$}_a-\b q)^2} d(\b{$\ell$}_a,\b q, \b
x_a)d(\b{$\ell$}_a,\b q, \b
x_a^\prime)\}$$
$$\cdot  \{-{\alpha^2\over 2} {d^2\ell_b\over \b{$\ell$}_b^2(\b{$\ell$}_b+ \b
q)^2}
d(\b
{$\ell$}_b,-\b  q, \b x_b) d(\b{$\ell$}_b,- \b q, \b x_b^\prime)\}\eqno(45)$$

\noindent where

$$d(\b{$\ell$}, \b q, \b x) = e^{i\b q\cdot \b x/2}+ e^{i\b q\cdot \b x/2}
-e^{i(\b
{$\ell$}-\b q/2)\cdot \b x} -e^{-i(\b{$\ell$}-\b q/2)\cdot \b x}  .\eqno(46)$$

\noindent The appearance of $d^2q$ in (45) is discussed in the Appendix.  The
two
$\{ \}$- terms in (45) give the dipole-dipole scattering as in (9) except for
the
nonzero momentum transfer in the present case.  Eq.(45) looks very complex
especially when one recalls that $n_2$ is given by (34) and (44).  However, in
the
leading logarithmic approximation there are important simplifications.  In the
leading logarithmic approximation  $x_a, x_b \cdot \cdot \cdot << 1/q$ and
$\ell_a,
\ell_b >> q.$  This means that we may set $\b q = 0$ in $d(\b{$\ell$}, \b q, \b
x)$,
drop $\b q$ in $(\b{$\ell$}_a-\b q)^2$ and in $(\b{$\ell$}_b-\b q)^2$, and use
the
asymptotic expression

$$e(\b x, \b q) \approx \ln({1\over qx}).\eqno(47)$$

\noindent Since  F,  as given by (45), is to be used in (8) we may take an
average
over the angles of $\b x_{01}$ and $\b x_{01}^\prime$.  Then the integrals in
(45)
factorize between the $a$  and  $b$  dipole scatterings, except for $a$
coupling
through the q-dependence.  Thus,

$$F={\pi^2(x_{01}x_{01}^\prime)^2 e^{2(\alpha_\rho-1)Y}\over 2(7 \alpha
C_F\zeta(3)Y)^6} \int {d\nu d\nu^\prime\over 4\pi^2} {V_\nu V_{\nu^\prime}\over
[4\ln 2-\chi(\nu)][4\ln 2-\chi(\nu^\prime)]}$$
$$\cdot \int {d^2q\over 4\pi^2} (q x_{01})^{-1+2i\nu} (q
x_{01}^\prime)^{-1+2i\nu^\prime}J^2\eqno(48)$$

\noindent where

$$J=4\pi\alpha^2 \int_0^\infty {dx dx^\prime\over (x x^\prime)^2} \int_q^\infty
{d\ell\over \ell^3} (1-J_0(\ell x))(1-J_0(\ell x^\prime)) \ln({1\over q
x})\ell
n({1\over q x^\prime})$$
$$\cdot exp\{-a(\ln^2 x_{01}/x + \ln^2
x_{01}^\prime/x^\prime)\}.\eqno(49)$$

\noindent It is straightforward to evaluate  J  in a manner similar to that
used
in evaluating  I  in sec.2.3.  We find

$$J =  \alpha^2(7 \alpha C_F \zeta(3)Y)^{3/2}\eqno(50)$$

\noindent where we have dropped nonleading terms in  Y.  Using (7) in (9) and
carrying out the q-integration yields

$$F={\alpha^4 e^{2(\alpha_\rho-1)Y} x_{01} x_{01}^\prime\over 16[7 \alpha
C_F\zeta(3)Y]^3} \int d\nu {(x_{01}/x_{01}^\prime)^{2i\nu}V_\nu V_{-\nu}\over
[4\ln 2 -\chi(\nu)]^2}.\eqno(51)$$

\noindent Eq.(51) along with(8) gives the two pomeron exchange contribution to
onium-onium scattering .  It would be very interesting to numerically compare
the
magnitudes of (26) and (51) to see at what values of Y two pomeron exchange
becomes
as important as one pomeron exchange.  Eq,(43) is easily inverted to give
$V_\nu$,
however, we have been unable to find a simple formula for $V_\nu.$

\bigskip
\centerline{\bf 4. The Triple Pomeron Coupling}

In this section we calculate the triple pomeron coupling.  As always our
calculation is carried out in the large $N_c$ limit.  Whether or not $1/N_c$
corrections to this quantity are calculable, in principle, is not clear[11].
Our
procedure of calculation is as follows.  (i)  We first calculate the dipole
pair
density in an onium state but where we require the high momentum part of the
longitudinal momentum evolution, between rapidities  Y  and $\bar y$, to be
given
by a
single pomeron while evolution between $\bar y$ and 0 be given by two pomerons.
(ii)
We then couple the two dipoles to separate onia states and use the resulting
expression to calculate large mass diffractive excitation, the process
traditionally
used to define the triple pomeron coupling.

Let $\bar n_2$ be the dipole pair correlation in an onium state where the two
dipoles
have transverse coordinate separations $\b x_a$ and $\b x_b$ while the smallest
light-cone momentum of a gluon making up part of either pair is $e^{-Y}p_+$
with
$p_+$ the onium momentum.  We further require that the longitudinal momentum
evolution from the scale $p_+$ to the scale $e^{(Y-\bar y)}p_+$ be that of
single
pomeron evolution while evolution below that scale be given by two independent
pomeron evolutions.  Then $\bar n_2$ obeys the equation

$$\bar n_2(Y,\bar y,\b x_{01}, \b x_a, \b x_b, \b q)
={2\alpha C_F\over \pi^2}\int{x_{01}^2d^2x_2\over x_{02}^2x_{12}^2}
exp\{-{4\alpha
C_F\over
\pi}
\ln(x_{01}/\rho)(Y-\bar y)\} e^{i\b q\cdot \b x_{01}/2}$$
$$n(\b x_{02}, \b x_a,\bar y, \b q)n(\b x_{12}, \b x_b,\bar y, \b q)+{2\alpha
C_F\over \pi^2}\int {x_{01}^2 d^2x_2\over x_{02}^2 x_{12}^2}$$
$$\int_{\bar y}^Y dy\
exp\{-{4\alpha C_F\over \pi}
\ln(x_{01}/\rho)(Y-y)\}
\bar n_2(y, \bar y, \b x_{12}, \b x_a, \b x_b, \b q)\eqno(52)$$

\noindent where, as before, q  is the transverse momentum carried by each of
the
dipoles.  Eq.(42) is illustrated in Fig.4.  Defining $n_2(Y,\bar y, x_{01}, q)$
exactly as in (34) one finds

$$\bar n_2(Y,\bar y, x_{01}, q) = {\alpha C_F e^{2(\alpha_P-1)\bar y}
x_{01}^2\over 2(7
\alpha C_F\zeta(3)\bar y)^3} exp\{-{4\alpha C_F\over \pi}
\ln(x_{01}/\rho)(Y-\bar y)\}$$
$$\int {d^2x_2\over x_{02}x_{12}} J_0(q
x_{01}/2) e(\b x_{12},\b q) e(\b x_{02}, \b q)
 + {4\alpha C_F\over \pi} \int
\tilde K(x_{01},x_{12}) dx_{12}$$
$$\int_{\bar y}^Y dy\  exp\{-{4\alpha C_F\over \pi}
\ln(x_{01}/\rho)(Y-y)\}\bar n_2(y,\bar y,x_{12}, q).\eqno(53)$$

\noindent  Going to the $\omega$-plane by defining

$$\bar n_2(Y,\bar y, x_{01}, q) = x_{01}^2\int {d\omega\over 2\pi}
e^{\omega(Y-\bar y)}(x_{01}q)^{-1+2i\nu}{d\nu\over 2\pi}
\bar n_{2\nu\omega}(\bar y)\eqno(54)$$

\noindent we obtain

$$\bar n_{2\nu\omega}(\bar y) = {\alpha C_FV_\nu e^{2(\alpha_P-1)\bar y}\over
2(7 \alpha
C_F\zeta(3)\bar y)^3(\omega -{4\alpha C_F\over \pi}\chi(\nu))}.\eqno(55)$$

\noindent From (55) it follows that

$$\bar n_2(Y,\bar y,x_{01}, q) = {\alpha C_FV_0x_{01}
e^{(\alpha_P-1)(Y-\bar y)+2(\alpha_P-1)\bar y}\over 8 q{\sqrt{7\alpha
C_F\zeta(3)(Y-\bar y)}}[7\alpha C_F\zeta(3)\bar y]^3}.\eqno(56)$$

\noindent The 3 onium $\to$ 3 onium amplitude is given by

$$A_6= - i\int d^2x_{01}d^2x d^2x^\prime\int_0^1dz_1dz dz^\prime
\Phi(x_{01},z_1)\Phi(x,z)\Phi(x^\prime,z^\prime)F_6\eqno(57)$$

\noindent with

$$F_6=\int \bar n_2(Y-\bar y/2,\bar y, \b x_{01}\b x_a,\b x_b,\b q)
 \{-{\alpha^2\over 2} {d^2\ell_a\over \b{$\ell$}_a^2(\b{$\ell$}_a-\b q)^2} d(\b
{$\ell$}_a, \ q, \b x_a) d(\b{$\ell$}_a,\b q, \b x_a^\prime)\}.$$
$$\cdot \{{-\alpha^2\over 2} {d^2\b{$\ell$}_b\over
\ell_b^2(\b{$\ell$}_b-q)^2}d(\b
{$\ell$}_b,
\b q,\b x_b)d(\b{$\ell$}_b, \b q, \b x_b^\prime)\}{d^2x_ad^2x_a^\prime\over
(2\pi
x_ax_a^\prime)^2}{d^2x_bd^2x_b^\prime\over (2\pi x_bx_b^\prime)^2}$$
$$\cdot n(\b x, \b x_a^\prime,\bar y/2,\b q) n(\b x^\prime,\b x_b^\prime, \bar
y/2,
\b
q)\eqno(58)$$

\noindent where $n(\b x,\b x^\prime, y, \b q)$ is given in (32).  ${\rm A}_6$
is
illustrated in Fig.5.  Eq.(58) is evaluated, using (56), in a manner almost
identical to that used in evaluating (45).  The result is

$$F_6={2\pi^2\alpha^5C_FV_0\over q} x_{01}x e(\b x, \b q) x^\prime e(\b
x^\prime,
\b q)
{e^{(\alpha_P-1)(Y-\bar y)}\over {\sqrt{7\alpha
C_F\zeta(3)(Y-\bar y)}}}{e^{2(\alpha_P-1)\bar y}\over (7\alpha
C_F\zeta(3)\bar y)^3}\eqno(59)$$

\noindent where ${\rm V}_0=V_\nu\vert_{\nu=0}$ with ${\rm V}_\nu$ given by
(43).
Unfortunatly, we have not been able to determine a definite value for $V_0$.

The triple pomeron coupling is clearly determined by ${\rm V}_0$.   However,
when
the pomeron is not a simple pole there is no unique way to normalize the triple
pomeron coupling.  To exhibit the relationship between $V_0$ and a physical
process
we need to relate $F_6$ to such a process.  One such way to do this would be to
use
$F_6$ to determine the diffractive part of the double scattering contribution
to
nuclear shadowing where, of course, our ``nucleus'' would be a nucleus made of
onia.  Perhaps a better process is the diffractive dissociation process onium
$(p)$ +
onium $(p^\prime) \to$ onium $(p^\prime-q) + X$ where $M^2=(p+q)^2$ is much
larger
than the onium mass squarred, but where $M^2$ is much less than
$s=(p+p^\prime)^2.$
In order to do this we need to assume the AGK cutting rules[12] which relate
the
value of $F_6$ to the discontinuity, in the variable $M^2$, producing
diffractive
dissociation.  The AGK rules say that ${1\over 2i}{\rm disc}_{M^2}A_6=-ImA_6.$
Thus,

$${d\sigma\over d\bar y} = - 2 Im A_6 {d^2q\over (2\pi)^2}\eqno(60)$$

\noindent The factor ${d^2q\over (2\pi)^2}$ occurs for reasons discussed in the
Appendix.  Using $d\bar y = {dM^2\over M^2}$ we obtain

$$M^2{d\sigma\over dtdM^2} = {e^{(\alpha_P-1)(Y-\bar y)}\over {\sqrt{7\alpha
C_F\zeta(3)(Y-\bar y)}}} {e^{2(\alpha_P-1)\bar y}\over (7\alpha C_F\zeta(3)\bar
y)^3}
2\alpha^5\pi C_F V_0{R\over {\sqrt{-t}}}\Phi^2(q)\eqno(61)$$

\noindent with  $q={\sqrt{-t}}$ and where

$$R = 2 \int d^2x \int_0^1 dz\  \Phi (x,z) x.\eqno(62)$$

\noindent Also,

$$\Phi (q) = \int d^2x \int_0^1 dz\  \Phi (x,z) x\  e(\b x, \b q).\eqno(63)$$

\noindent For reference, we note that for elastic onium-onium scattering

$${d\sigma\over dt} = {e^{2(\alpha_P-1)Y}\over (7\alpha C_F\zeta(3)Y)^3}
\alpha^4
\Phi^4(q).\eqno(64)$$

The $1/{\sqrt{-t}}$ factor in (61) is perhaps surprising.  Such a  factor means
that the triple pomeron coupling is singular at t=0.  There is also a mild,
logarithmic, singularity in $\Phi(q)$ as $q\to 0$.  It is also perhaps
worthwhile
to note that in writing (53) we have assumed that the $x_a$ and $x_b$
dependence of
$\bar n_2(Y,\bar y,\b x_{01}, \b x, \b q)$ is given by an equation essentially
identical
to (34).  This may appear an unnatural assumption, especially if $Y-\bar y
\approx
\bar y/2$ since one might expect that the transverse coordinate separation at
the
point
of joining of the three pomerons has {\it diffused} far from the size $x_{01}$
characterizing the onium state  p.  If $x_c$ is the transverse coordinate
separation
characterizing the point of joining of three pomeron we might expect $\ln^2
x_c/x_{01} \sim {14\alpha C_F\zeta(3)(Y-\bar y)\over \pi}[2].$   However,
because
of
the nonzero q it is necessary that $x_c \leq 1/q$ while short distances are
suppressed at the connection point, as is apparent from (52).  Thus, $x_c$ is
determined by the size 1/q.  In order that (53) be correct we need to require
that
$\vert \ln^2 x_{01}/x_a - \ln^2 1/q x_a \vert << a(\bar y)^{-1}$ thus
allowing
the exponential part of the $x_a$ dependence to be given in terms of $x_{01}$
as
in (34).  This requires

$$\ln^2 q x_{01} << {14\alpha C_F\zeta(3)\bar y\over \pi}\eqno(65)$$

\noindent to be satisfied when using (59) or (61).
\bigskip
\bigskip
\centerline{\bf Appendix}

In sec.2, we have viewed high energy onium-onium scattering in terms of the
lowest
order color neutral exchange of two gluons between left-moving and right-moving
dipoles found in the colliding onium states.  This is expressed in (20).  Our
purpose in this section is to derive the elementary dipole scattering
amplitude,
(9), and provide the necessary generalizations for sections 3 and 4.

We begin by referring to Fig.1 where two gluons are exchanged between two onium
states.  In the center of mass system and at high energies the gluons are
exchanged over a period of time short compared to the, dilated, interaction
times
within the individual onium states.  Thus, during the actual time the gluons
are
being exchanged one can view the onia simply as two free quark-antiquark pairs,
or
color dipoles.  The calculation leading to (7) and (9) can be done in terms of
the
scattering of a free quark-antiquark pair (dipole) on a free quark-antiquark
pair
(dipole).  The three classes of graphs which occur are illustrated in Fig.6,
where
we label momenta in such a way that the two dipoles do not exchange momenta.
The
contribution of the graph shown in Fig.6a is

$$G_a={g^4/4\over
2k_{1+}2k_{1-}^\prime
}\int[\tilde u(k_1)\gamma_\mu\gamma\cdot(k_1+\ell)\gamma_\nu u(k_1)]
[\tilde u(k_1)\gamma_\mu\gamma\cdot(k_1^\prime -\ell)\gamma_\nu
u(k_1^\prime)]{d^4\ell\over (2\pi)^4}$$
$${(2\pi)^3\delta^3(k_1-\bar k_1)(2\pi)^3\delta^3(k_1^\prime-\bar k_1^\prime
)\over
[(k_1+\ell)^2+i\epsilon][(k_1^\prime-\ell)^2+i\epsilon][\ell^2+i\epsilon]^2}
{d^3k_1\over (2\pi)^3} {d^3\bar k_1\over (2\pi)^3} {d^3k_1^\prime\over
(2\pi)^3} {d^3\bar k_1^\prime\over (2\pi)^3} + \cdot \cdot \cdot\eqno(A.1)$$

\noindent where the $\delta$-functions refer to the disconnected lines  and
$\delta^3(k_1-\bar k_1) = \delta(k_{1+}-\bar k_{1+})\delta(\b k_1-\bar {\b
k}_1)$ while
$\delta^3(k_1^\prime-\bar k_1^\prime) = \delta(k_{1-}^\prime -\bar
k_{1-})\delta(\b
k_1^\prime-\bar  {\b k}_1^\prime).$  That is, for the unprimed dipole $\b k$
and
$k_+$
are the momenta labelling these right-movers while for the primed dipole $\b
k_1^\prime$  and  $k_-^\prime$ are the momenta labelling these left-movers.
Adding
in the crossed graph to the expression in (A.1) one finds

$$G_a = - {\alpha^2\over 2}\int {d^2\ell\over [\b{$\ell$}^2]^2} {d^3k_1\over
(2\pi)^3}\ {d^3k_1^\prime\over (2\pi)^3}\eqno(A.2)$$

\noindent as in (7) except for the extra phase space factors.

Now turn to the graph shown in Fig.6c for which one has the expression

$$G_c = -  g^4/4[\tilde u(p-\bar k_1)\gamma_\mu u(p-k_1)][\tilde
v(k_1^\prime)\gamma_\nu
v(\bar k_1^\prime)]$$
$$\cdot [\tilde u(p^\prime-\bar k_1^\prime)\gamma_\mu
u(p_1^\prime-k_1^\prime)][\tilde v(k_1^\prime)\gamma_\nu v(\bar k_1^\prime)]$$
$$[2k_{1+}2(p-k_1)_+2k_{1-}^\prime
2(p_1^\prime-k_1^\prime)_-[(k_1-\bar
k_1)^2+i\epsilon]^2]^{-1}(2\pi)^4\delta^4(k_1+k_1^\prime-\bar k_1-\bar
k_1^\prime)$$
$$\cdot {d^3k_1\over (2\pi)^3} {d^3\bar k_1\over (2\pi)^3} {d^3k_1^\prime\over
(2\pi)^3} {d^3\bar k_1^\prime\over (2\pi)^3}.\eqno(A.3)$$

\noindent One can integrate $d^3\bar k_1^\prime d\bar k_{1+}$ to eliminate the
$\delta$-function in (A.3).  Define $\b{$\ell$} = \bar {\b k}_1-\b k_1.$  Then

$$G_c= - \alpha^2 \int {d^2\ell\over [\b{$\ell$}^2]^2}\ {d^3k_1\over (2\pi)^3}\
{d^3k_1^\prime\over (2\pi)^3}.\eqno(A.4)$$

When going to transverse coordinate space, with $\b x_{01}$ congugate to $\b
k_1$, and $\b x_{01}^\prime$ conjugate to $\b k_1^\prime$, (A.2) corresponds to
the
term obtained by taking the first term in each of the factors on the right-hand
side
of (9).  The factor of 4 counts the number of one-loop contributions involving
only a
pair of partons.  The term (A.4) corresponds to terms like
$e^{-i\b{$\ell$}\cdot
\b
x_{01} +i\b{$\ell$}\cdot \b x^\prime_{01}}$  and $e^{i\b{$\ell$}\cdot \b
x_{01}-i\b
{$\ell$}\cdot \b  x_{01}^\prime}$
in (9).  Factors having
$e^{-i\b{$\ell$}\cdot\b x_{01}-i\b{$\ell$}\cdot\b x_{01}^\prime}$ and
$e^{i\b
{$\ell$}\cdot
\b x_{01} + i\b{$\ell$} \cdot \b x_{01}^\prime}$
come from terms where the
quark in  p  interacts with an antiquark in ${\rm p}^\prime$ and the antiquark
in  p
interacts with a quark in ${\rm p}^\prime.$  Graphs of the types shown in
Fig.6b
account for factors $e^{-i\b{$\ell$}\cdot \b x_{01}}$ etc.

The crucial point to emerge from this analysis is that only a factor
${d^2\ell\over (2\pi)^2}$ remains from the factor ${d^4\ell\over (2\pi)^4}$
which
appears in writing down Feynman diagrams.  The reduction of ${d^4\ell\over
(2\pi)^4}$ to ${d^2\ell\over (2\pi)^2}$ occurs in a different manner for the
different contributions.  For the contribution shown in Fig.6a contour
distortions
over the two fermion poles in the one-loop graph eliminate $d\ell_+ d\ell_-$
while
for graphs like those in Fig.6c energy momentum $\delta$-functions eliminate
the
longitudinal integrations.

When two pairs of gluons are exchanged  between pairs of dipoles in each of two
colliding onia the same result holds for each and every exchanged momentum. For
example, in the circumstance discussed in sec.3 where  q  is the momentum
transfer from a single dipole in one onium to a single dipole in the second
onium
state by a pair of gluons, the disconnected nature of the gluon exchanges
eliminates
${dq_+dq_-\over (2\pi)^2}$ exactly as for the contribution $G_c$ above.  Thus
the
factor ${d^2q\over (2\pi)^2}$ naturally appears in (45).

\bigskip
\centerline{\bf References}
\item{[1]}			Ya. Ya. Balitsky and L.N. Lipatov, Sov.J. Nucl. Phys.28 (1978)
822.
\item{[2]}			E.A. Kuraev, L.N. Lipatov and V.S. Fadin, Sov. Phys. JETP 45
(1977)
199.
\item{[3]}			L.N. Lipatov, Sov.Phys. JETP 63 (1986) 904.
\item{[4]}			A.H. Mueller and H. Navelet, Nucl. Phys. B282 (1987) 727.
\item{[5]}			W.-K. Tang, Phys. Lett. 278B (1992) 363.
\item{[6]}			J. Bartels, J. De Roeck and M. Loewe, Z. Phys. C54 (1992) 635.
\item{[7]}			J. Kwiecinski, A. Martin and P.J. Sutton, Phys. Lett.287B (1992)
254;
phys. Rev.D46 (1992) 921.
\item{[8]}			L.L. Frankfurt and M.I. Strikman, Phys. Rev. Lett. 63 (1989) 1914.
\item{[9]}			A.H. Mueller, CU-TP-609 (August 1993). (To be published in Nuclear
Physics B).
\item{[10]}		B. Andersson, G. Gustafson, A. Nilsson and C.
Sj\"ogren, Z. Phys. C49 (1991) 79.
\item{[11]}		L.N. Lipatov in ``Perturbative Quantum Chromodynamics,"
ed.,A.H.\break Mueller, World Scientific, Singapore 1989.
\item{[12]}		V.A. Abramovskii, V.N. Gribov and O.V. Kancheli, Sov.J. Nucl.
Phys. 18
(1974) 308.

\end
\bye